\documentclass{JHEP3}
\usepackage{epsfig}
\usepackage{graphicx}% Include figure files
\usepackage{dcolumn}% Align table columns on decimal point
\usepackage{bm}% bold math
\newcommand{\be}{\begin{equation}}
\newcommand{\ee}{\end{equation}}
\def\bea{\begin{eqnarray}}
\def\eea{\end{eqnarray}}

\usepackage{amssymb}
\usepackage{amsmath}
\usepackage{amsfonts}
 \def\be{\begin{equation}}
\def\ee{\end{equation}}
\def\bea{\begin{eqnarray}}
\def\eea{\end{eqnarray}}
\def\lesssim{\mathrel{\hbox{\rlap{\hbox{\lower4pt\hbox{$\sim$}}}\hbox{$<$}}}}
\def\gtrsim{\mathrel{\hbox{\rlap{\hbox{\lower4pt\hbox{$\sim$}}}\hbox{$>$}}}}

\title{From BPS to Non-BPS
Black Holes Canonically}
  \author{Renata Kallosh \footnote{\mbox{
Email: {\tt kallosh@stanford.edu} } }
\\
Department of Physics, Stanford University, Stanford, CA 94305,
USA}
\received{\today} %%
 \preprint{SU-ITP-06/06 
 \\ February 28 2006 }
\abstract{ We describe the  ``action-angle'' integrable  system underlying the structure of double-extremal  
black holes. This implies the existence of a canonical transformation from BPS to 
non-BPS black  holes.  We give examples of such canonical transformation 
for  STU and for $E_{7(7)}$-invariant black holes.
}
\begin{document}

\section{Introduction}

During the last 15 years we have learned how to find extreme black hole solutions with unbroken supersymmetry by solving equation ${\cal D}Z=0$, where $Z$ is the vector multiplet central charge in N=2 supergravity.  The BPS solutions satisfy the near-horizon attractor equations \cite{Ferrara:1995ih} which have been studied for a long time.  This  is much simpler than to solve directly  a system of equations for all fields, including the Einstein equations and take the near-horizon limit afterwards.  Various relations between BPS near-horizon geometries have been established. For non-supersymmetric black holes these methods do not directly apply because they do not satisfy the simple equation ${\cal D}Z=0$.  The non-BPS attractors \cite{Ferrara:1997tw}-\cite{Bellucci:2006ew}  are much less studied and only recently the relevant non-BPS attractor equations \cite{Kallosh:2005ax}-\cite{Bellucci:2006ew}   have been derived. As a result, non-supersymmetric black holes are much less known and understood than the supersymmetric ones. 

The purpose of this note is to describe the relation between extremal black holes with unbroken supersymmetry, satisfying
 equation ${\cal D}Z=0$, and extremal black holes with spontaneously broken supersymmetry, with ${\cal D}Z\neq 0$.   
We will be able  to generate  the non-BPS solutions from the BPS ones in a canonical way. 
For this purpose we will first remind some well known facts about the one-dimensional action related to extremal black holes in four dimensions. We will then formulate an integrable system relevant to double-extremal black holes with everywhere constant scalars satisfying the BPS/non-BPS attractor equation. We will explain the canonical structure of this system and the reason for the existence of a canonical transformation between BPS and non-BPS attractors. We exemplify  these general arguments by few simple examples.

\section{Extremal black holes and attractors} 

We start with  the
bosonic part of the Einstein-Maxwell action coupled to some Abelian
vector fields, as in  \cite{Ferrara:1997tw}-\cite{Kallosh:2006bt} 
\begin{equation}
-{R\over 2} + G_{a\bar a} \partial_ \mu z^a   \partial_\nu \bar
z^{\bar a} g^{\mu\nu} +
 {\rm Im} {\cal N}_{\Lambda \Sigma} {\cal F}^{\Lambda}_{\mu \nu}  {\cal
F}^{ \Sigma}_{\lambda \rho}  g^{\mu \lambda} g^{\nu \rho} + {\rm Re}
{\cal N}_{\Lambda \Sigma} {\cal F}^{\Lambda}_{\mu
\nu}\left(\ast{\cal F}^{ \Sigma}_{\lambda \rho}\right) g^{\mu
\lambda} g^{\nu \rho}
 \ .
\label{scalaraction2}
\end{equation}
We use the static spherically symmetric ansatz for the extremal black hole
 metric
\begin{equation}
ds^2 = e^{2U} dt^2 - e^{-2U} \left [{ d
\tau^2 \over \tau^4}  + {d \Omega^2 \over  \tau^2} \right]\ . \label
{ansatz}
\end{equation}
 Taking into account the independence of the action from the electric
$\psi^\Lambda(\tau)$ and magnetic $\chi_\Lambda(\tau)$
potentials due to
gauge invariance of the action, one finds the
one-dimensional Lagrangian for the evolution of $U(\tau), z(\tau)$
and $\bar z(\tau)$   \cite{Ferrara:1997tw},
\begin{equation}
 {\cal L} \left (U(\tau) , z^a(\tau) , \bar z^{\bar a}(\tau) \right )= \left ({d U
\over d\tau}\right )^2 +  G_{a\bar a} {dz^a \over d \tau} {d \bar
z^{\bar a} \over d \tau}
 + e^{2U} V_{BH}(z,\bar z, p,q)
\ . \label{lagr}
\end{equation}
with the  constraint:
\begin{equation}
  \left ({d U
\over d\tau}\right )^2 +    G_{a\bar a} {dz^a \over d \tau} {d \bar
z^{\bar a} \over d \tau}
 - e^{2U} V_{BH}(z,\bar z, p,q) =0
\ . \label{constr1}
\end{equation}
Here the ``black hole potential''  \cite{Ferrara:1997tw}, \cite{Goldstein:2005hq}, \cite{Tripathy:2005qp} is:
\begin{equation}
V_{BH}(z,\bar z, p,q)= |Z(z,\bar z, p,q)|^2 + |{\cal D}_a
Z(z,\bar z, p,q)|^2 )\ , \label{pot1}
\end{equation}
$Z$ is the central charge, the charge of the graviphoton in ${\cal
N}=2$ supergravity and ${\cal D}_a Z$ is the K\"{a}hler covariant
derivative of the central charge.
\begin{equation}
Z(z, \bar z, q,p) = e^{K(z, \bar z)\over 2} (X^\Lambda(z)  q_\Lambda
- F_\Lambda(z) \, p^\Lambda)= (L^\Lambda q_\Lambda - M_\Lambda
p^\Lambda) \ . \label{central}
\end{equation}
For double-extreme black holes \cite{Kallosh:1996tf}-\cite{Behrndt:1996jn} scalars are constant: their values at infinity, $z_{\infty},\bar z_{\infty}$ are the same as near the horizon, $z_{fix}, \bar z_{fix}$. 

The attractor equation for non-BPS and BPS black holes was proposed in \cite{Kallosh:2005ax} and developed in \cite{Kallosh:2006bt}.  It defines the fixed values of the moduli $z_{fix}, \bar z_{fix}$ as the functions of the black holes charges, $(p, q)$.
\begin{eqnarray}\label{attr}
H_3=2\mbox{Im}\left [Z\, \overline { \Omega}_3- \frac{(\overline
{\mathcal{D}}_{\bar a}\overline {\mathcal{D}}_{\bar b} \overline Z)
G^{\bar{a} a}G^{\bar b b }{\mathcal{D}}_{{b}}{Z} }{2{Z}}
{{\mathcal{D}}}_{{a} } { \Omega}_3\right] \qquad \Rightarrow \qquad z_{fix}(p,q), \quad \bar z_{fix}(p,q) \ .
\end{eqnarray}
In case of BPS black holes the second term in the right hand side of eq. (\ref{attr}) vanishes and it is reduced to the BPS black hole attractor equation \cite{Ferrara:1995ih}.

At infinity, as $\tau \rightarrow 0$,
$U\rightarrow M\tau$ and one finds a Minkowski metric and the
constraint:
\begin{equation}
M ^2 ( p,q ) =  |Z(z_{fix} ,\bar
z_{fix}, p,q)|^2  + |{\cal D}_a Z(z_{fix} ,\bar
z_{fix},
p,q)|^2 \ .
\label{M2}\end{equation}
 The BPS configuration has its
mass equal to the central charge in supersymmetric theories so that:
\begin{equation}
M ^2 ( p,q ) =    |Z(z_{fix}, \bar z_{fix}, p,q)|^2   \ ,   \qquad   
{\cal D}_a Z(z_{fix}, \bar z_{fix}, p,q )  =0 \ .\label{BPS}
\end{equation}
The non-BPS extremal black holes have a mass squared given in expression (\ref{M2}). The double-extremal non-BPS black holes were presented in \cite{Kallosh:2006bt}. In both cases, BPS and non-BPS,  the double-extremal black hole entropy is given by the following expression
\be
S(p,q)= \pi M ^2 ( p,q ) \ .
\ee
At the attractor point the mass formula can be presented in terms of a symplectic invariant 
 $I_1$ of the special geometry
 constructed in \cite{Ceresole:1995ca}
\begin{equation}
M^2(p,q)=I_1 (p,q) = |Z(z,\bar z, p,q)|^2 + |{\cal D}_i Z(z,\bar z, p,q)|^2=- {1\over 2}(p,q) \cdot {\cal M}(z_{fix},  \bar z_{fix})\cdot
  (p, q)^t
\label{M2}\end{equation}
where the matrix ${\cal M}(z, \bar z)$ at the arbitrary point of moduli space is given by
\begin{equation}
{\cal M}({\cal N})=   \left |\begin{array}{cc}
{\rm Im} {\cal N} + {\rm Re} {\cal N}  {\rm Im} {\cal N} ^{-1} {\rm Re} {\cal
N}  &-{\rm Re} {\cal N}  {\rm Im} {\cal N}^{-1}  \cr-
 {\rm Im} {\cal N}^{-1} {\rm Re} {\cal N} &  {\rm Im} {\cal N}^{-1} \cr
\end{array}\right |
\label{calM}\end{equation}
and ${\cal N}$ depends on special coordinates $(z, \bar z)$ and ${\rm Im} {\cal N}$ is negative definite.

\section{Double-extremal black holes as an integrable system}

\subsection{Kamiltonian}

Consider a  canonical system with some Hamiltonian which can be transformed into a form in which 
the Kamiltonian $K(J_i)$ (the Hamiltonian after the transformation to action-angle variables) of the integrable system depends only on the  action variables $J_i$ and does not depend on angle variables $\Phi^i$.
The Kamiltonian equations of motion for action variables $J_i$ and angle variables $\Phi^i$ are
\bea\label{Q}
 \dot J_i = -{\partial K\over \partial  \Phi^i}=0 \label{P}\ , \qquad  \dot \Phi^i ={\partial K\over \partial  J_i} = {\cal V}^i (J_k) \qquad \Rightarrow \qquad
 \Phi^i = \Phi^i_0 +  {\cal V}^i (J_k)\,  \xi \ .
\eea
The first equation  states that all our $J_i$-variables do not depend on the evolution parameter $\xi$. The second equation allows to calculate the frequencies ${\cal V}^i(J)$ as the functions of action variables.

 One can, in general, perform a Legendre transform to the frequency variables ${\cal V}$, instead of the action variables $J$.
\be
K(J)= \tilde K({\cal V}) + J_i {\cal V}^i 
\label{Legendre}\ee
where
\be
{\partial K(J)\over \partial  J_i} = {\cal V}^i  \ , \qquad {\partial \tilde K({\cal V})\over \partial {\cal V}^i} = -J_i 
\ee

\subsection{Double-extremal black holes}

We can consider the effective one-dimensional theory of double-extremal black holes, with scalars fixed everywhere to attractor values,  as a case of integrable system. At this point we have to change the evolution parameter $\tau$ to $\xi$:
\be
\xi = {\tau \over 1-M\tau}\ ,  \qquad \tau = {\xi\over 1+M\xi} \ .
\ee
The reason for the change of variables is the following: equations of motion for the electric and magnetic potentials are \cite{Kallosh:1996tf,Kallosh:2006bt}, 
\be
e^{-2U} \partial_\tau (\psi, \chi)= -  {\cal M}\cdot (p,q)^t \ ,
\ee
where for the double-extremal black holes 
\be
e^{-U}= 1- M \tau = {1\over 1+M\xi} \ .
\ee
Therefore the equation for electric-magnetic potentials can be presented in the form
\be
 \partial_\xi (\psi, \chi)= -  {\cal M}\cdot (p,q)^t \ .
\ee
We have a set of $\xi$-independent action variables 
$
J= (J_M, p , q )$
 and a set of 
angle  variables which depend on $\xi$ linearly:
\be
\Phi^i= \left (e^{U}(\xi), \psi (\xi), \chi(\xi) \right)\ ,
\ee 
\bea \label {M}
e^{U}(\xi) &=& 1+M \xi \ ,\\
 (\psi (\xi), \chi (\xi))   &=&  (\psi_0, \chi_0) - {\cal M} \cdot ( p, q)^t  \, \xi \ .
\label {psi}\eea
The scalars are fixed by attractor equations and are functions of $(p,q)$, $z_{fix}(p,q), \bar z_{fix} (p,q)$.
However, not all of these variables are independent, since we have a constrained system: the pair $(J_M, e^U)$ can be removed. We therefore find that 
the frequency $M$ in the expression in (\ref{M})  can be replaced by the function of only $(p,q)$ as follows
\be \label {M1}
e^{U}(\xi) = 1+M \xi \ , \qquad M=M(p,q)= \left|-{1\over 2}(p,q)\cdot {\cal M }\left (z_{fix}(p,q), \bar z_{fix}(p,q) \right ) \cdot (p,q)^t\right| ^{1/2} 
\ee
The Kamiltonian  of the integrable system depending only on the independent action variables $(p, q)$ can be presented as
\be
K= M^2 (p, q)  \ ,
\label{K}\ee
where $M^2 (p, q)$ is defined in eqs. (\ref{M2}), (\ref{calM}).

To prove our assertion we have to calculate the right hand side of eq. (\ref{Q}) and compare it with eq.  (\ref{psi}). We find
\be
{\partial K\over \partial  (p,q)}= - {\cal M }\left (z_{fix}(p,q), \bar z_{fix}(p,q) \right ) \cdot (p,q)^t \ .
\ee
Note that we have used the fact that 
\be
{\partial \over \partial z} \left ((p,q)\cdot {\cal M }\left (z, \bar z \right ) \cdot (p,q)^t\right)=0 \ ,
\ee
since this is the critical point of the potential, ${\partial \over \partial z} V_{BH}=0$.

Thus we see that any double-extremal black hole solution corresponds to a canonical integrable system. At the attractor point the canonical transformations are $Sp(2(n+1), \mathbb{Z})$ transformations, in models with $n+1$ electric and $n+1$ magnetic charges.

\subsection{Legendre transform and Hesse potential}

We may perform a Legendre transform on our Kamiltonian (\ref{K}) and find the relation to the  Hesse potential, discussed recently in 
\cite{LopesCardoso:2006bg}. We perform the  Legendre transform from the action variables $J_i=(p^\Lambda, q_\Lambda)$ to the frequencies ${\cal V}^i(J)=(\dot \psi^\Lambda, \dot \chi_\Lambda) $ which are proportional to $(x^\Lambda, y_\Lambda)$  in \cite{LopesCardoso:2006bg}. To simplify the relation to \cite{LopesCardoso:2006bg} we will use $(\dot \psi^\Lambda, \dot \chi_\Lambda)= 2 ( x^\Lambda, y_\Lambda)$. The general case of Legendre transform on  Kamiltonian $K(J)$ in eq. (\ref{Legendre}) takes the form
\be
K(p,q) = 2 [ {\cal H}(x, y)+ p^\Lambda y_\Lambda - q_\Lambda x^\Lambda]
\label{Leg}\ee
Here 
\be
\tilde K (x, y)= 2 {\cal H}(x, y)=  (x,y) \cdot {\cal M}^{-1}\Big (z_{fix}(x,y),  \bar z_{fix}(x,y)\Big)\cdot
  (x,y)^t \ ,
\label{calH} \ee
${\cal M}^{-1}$ is the matrix inverse to ${\cal M}({\cal N})$ defined in eq. (\ref{calM})
and
\be
z_{fix}(x,y) = z_{fix}\Big (p(x,y), q(x,y)\Big ) \qquad \bar z_{fix}(x,y) = \bar z_{fix}\Big (p(x,y), q(x,y)\Big ) 
\ee
The derivative of the Kamiltonian over $(p,q)$ is given by
\be
{\partial K(p,q)\over \partial (p,q)}= -{\cal M}\Big (z_{fix}(p,q),  \bar z_{fix}(p,q)\Big ) \cdot (p,q)^t=
(\dot \psi, \dot \chi)^t= 2 (x,y)^t
\label{fr}\ee
and
\be
{\partial {\cal H}(x, y)\over \partial x^\Lambda}= q_{\Lambda} \ , \qquad {\partial {\cal H}(x, y)\over \partial y_\Lambda}=- p^{\Lambda}
\ee
To compare with \cite{LopesCardoso:2006bg} we have to keep in mind  that at this stage, when considering BPS and non-BPS configurations,  the higher derivative terms have not yet been included. If, in addition, we would for the sake of comparison consider only the BPS case with $DZ=0$, we would be able to use the fact that 
\begin{equation}
{\cal M}({\cal F})=   \left |\begin{array}{cc}
{\rm Im} {\cal F} + {\rm Re} {\cal F}  {\rm Im} {\cal F} ^{-1} {\rm Re} {\cal
F}  &-{\rm Re} {\cal F}  {\rm Im} {\cal F}^{-1}  \cr-
 {\rm Im} {\cal F}^{-1} {\rm Re} {\cal F} &  {\rm Im} {\cal F}^{-1} \cr
\end{array}\right | \ , \qquad DZ=0
\end{equation}
where ${\cal F}_{IJ}\equiv \partial_I \partial_J {\cal F}$ and ${\cal F}$ is the prepotential. In such case, our eq. (\ref{Leg}) is reduced to the Legendre transform between the BPS black hole entropy $S(p,q)$ and the Hesse potential ${\cal H}(x,y)$ proposed in \cite{LopesCardoso:2006bg}. In more general case of either non-BPS black holes or non-existence of the prepotential, the relevant matrix ${\cal M}(z, \bar z)$ is defined in eq. (\ref{calM}) and the matrix ${\cal N}$ can be constructed from the symplectic section $(X^\Lambda, F_{\Lambda})$ as explained in \cite{Ceresole:1995ca}. The Legendre transform (\ref{Leg}) remains valid with the definitions given in eq. (\ref{calH}).

\section{Canonical transform between BPS and non-BPS black holes}
There are 2 invariants under symplectic transformations \cite{Ceresole:1995ca}:
\bea
I_1(p,q; z, \bar z)&=& |Z|^2+|{\cal D}Z|^2 = -{1\over 2} (p,q)\cdot \Big ({\cal M}({\cal N})\Big )\cdot (p,q)^t\ , \\
I_2(p,q; z, \bar z) &=&  |Z|^2-|{\cal D}Z|^2= -{1\over 2} (p,q)\cdot \Big ({\cal M}({\cal F})\Big ) \cdot(p,q)^t \ .
\eea
The first invariant is always present as long as special geometry is defined by a choice of a symplectic section $(X^\Lambda, F_{\Lambda})$. The existence of the second invariant requires that the prepotential exists. This may or may not be the case, as shown in \cite{Ceresole:1995jg}. In such case, we may expect that the canonical $Sp(2(n+1), \mathbb{Z})$  transformations under certain conditions may not preserve $I_2$. Since they do preserve $I_1$, this means that neither $|Z|^2$ nor $|{\cal D}Z|^2$ will be preserved. Since we know examples of such BPS and non-BPS black holes \cite{Tripathy:2005qp}, \cite{Kallosh:2006bt}, it would be interesting to find out the relevant canonical transformation which does not preserve the value of ${\cal D} Z$.
\be
{\cal D} Z=0 \qquad \Leftrightarrow \qquad {\cal D} Z\neq 0 \ .
\ee
Both of these black holes near the horizon satisfy the new attractor equations, \cite{Kallosh:2005ax}, \cite{Kallosh:2006bt}. It is useful to present them in the form proposed in \cite{Bellucci:2006ew}.
\be
(Y_{ij} h^j)_{\partial V=0}=0 \ ,
\ee
where $h^j \equiv g^{j\bar j} \overline {\cal D}_{\bar j} \overline Z$ and $Y_{ij}\equiv {\cal D}_i {\cal D}_j Z +2{\overline Z\over |{\cal D} Z|^2} {\cal D}_i Z {\cal D}_j Z$.  There are two types of  solution of this equation : 
 BPS, $h^i=0$
and  non-BPS, $h^i\neq 0$, $\det \, Y=0$.
Clearly, it is easy to find BPS solutions with ${\cal D} Z=0$. How can we generate the non-BPS solution from it? A generic symplectic transformation at the attractor point (or for the double-extreme black holes)  is given by
\be \label{can}
\begin{pmatrix}
  p \\
 q 
\end{pmatrix}'= 
\begin{pmatrix}
  A & B \\
  C & D 
\end{pmatrix}
\begin{pmatrix}
  p \\
q
\end{pmatrix}\ , \qquad \begin{pmatrix}
  A & B \\
  C & D 
\end{pmatrix} \subset Sp(2(n+1), \mathbb{Z})\ .
\ee
The black hole entropy $S= \pi(|Z|^2+|{\cal D} Z|^2)_{fix}$ is invariant under these transformations. The values of scalars are defined by the attractor equations which for the BPS black holes are known in various cases. The general structure of these transformations in the canonical system described before implies that they include the interesting ones, transforming BPS into non-BPS.

\subsection {STU black holes}
Consider the STU black holes
\cite{Behrndt:1996hu} in the basis where there is no prepotential. The solution of the BPS attractor equations gives the following values for the entropy in terms of 4 magnetic and 4 electric charges $\hat p^\Lambda, \hat q_{\Lambda}$.
\be
{S\over \pi} =M^2 = \sqrt {\hat p^2 \hat q^2 -\hat p \cdot \hat q}\ , \qquad p^2 \hat q^2 -\hat p \cdot \hat q>0 \ .
\ee
Here $\hat p^2= \hat p^\Lambda \eta_{\Lambda \Sigma} \hat p^\Sigma$, etc. where the $SO(2,2)$-invariant metric is $++--$. The position of the vectors can be uplifted using the $SO(2,2)$-invariant metric, e. g. $\hat q^\Lambda = \eta^{\Lambda \Sigma} \hat q_\Sigma$.
The attractor values of the moduli are 
\begin{equation}
S= {\hat p \cdot \hat q   - i \left (\hat p^2 \hat q^2 - (\hat p
\cdot \hat
q)^2\right )^{1/2} \over \hat p^2} \ ,
\end{equation}
\begin{equation}
T=  {\bar S (\hat
p^3 - \hat
p^1) - (\hat q^3 - \hat q^1) \over \bar S (\hat p^0 - \hat p^2) -
(\hat q^0 -
\hat q^2) } \ ,
\end{equation}
\begin{equation}
U=  {\bar S (-\hat
p^3 -
\hat p^1) - (-\hat q^3 - \hat q^1) \over \bar S (\hat p^0 - \hat p^2)
- (\hat
q^0 - \hat q^2) } \ ,
\end{equation}
and the BPS mass/area formula is
\begin{equation}
Z \bar Z = M^2 = {A\over 4\pi} = \left (\hat p^2 \hat q^2 - (\hat p
\cdot \hat
q)^2\right )^{1/2} \ .
\end{equation}
To transform this solution to the non-BPS one we first have to take a simplified solution which has only 4 non-vanishing charges, out of 8. These are $\hat p^1, \hat p^3, \hat q_0, \hat q_2$. Out simplified solution has no axions,
\begin{equation}
S= {   - i \left (\hat p^2 \hat q^2 \right )^{1/2} \over \hat p^2} \ ,
\qquad
T=  {\bar S (\hat
p^3 - \hat
p^1)  \over  -
(\hat q^0 -
\hat q^2) } \ ,
\qquad 
U=  {\bar S (-\hat
p^3 -
\hat p^1)  \over 
- (\hat
q^0 - \hat q^2) } \ ,
\label{BPS}\end{equation}
and the mass/area formula is
\begin{equation}
Z \bar Z = M^2 = {A\over 4\pi} = \left (\hat p^2 \hat q^2 \right )^{1/2} \ .
\end{equation} 
Consider the following canonical transformation in eq. (\ref{can}) where 
\be\label{can1}
A= D=
\begin{pmatrix}
  ~0~ & ~0~  & ~1~ & ~0~ \\
  0 & 1 & 0 & 0 \\
  1 & 0 & 0 & 0 \\
  0 & 0 & 0 & 1 
\end{pmatrix} \ , \qquad B=C=0 \ .
\ee
The charge vector becomes 
$(0, \hat p^1, 0, \hat p^3, \hat q_0, 0, \hat q_2, 0)'= (0, \hat p^1, 0, \hat p^3, \hat q_2, 0, \hat q_0, 0)$.
Thus, we find that
\be
\hat p^2 = (\hat p')^2 \ , \qquad \hat q^2 = -(\hat q')^2 \ , \qquad  \hat p \cdot \hat q = (\hat p \cdot \hat q)'=0 \ .
\ee
Since the transformation (\ref{can}), (\ref{can1}) is canonical, it must preserve the positive invariant $I_1$ and the entropy/area of the black hole
\be
{S\over \pi} = {A\over 4 \pi}= |Z|^2+|{\cal D} Z|^2  = 
\sqrt {\hat p^2 \hat q^2}
 = |Z'|^2+|{\cal D} Z  '|^2  = 
\sqrt {-(\hat p')^2   (\hat q')^2 } \ .
\ee
The original system with $\hat p, \hat q$ charges and fixed moduli in eq. (\ref{BPS}) was for BPS black holes, the one with primed charges $\hat p', \hat q'$ is for non-BPS black holes. The attractor values of moduli of the non-BPS black holes are given by the expressions
\begin{equation}
S'= {   - i  \sqrt {-(\hat p')^2   (\hat q')^2 } \over (\hat p')^2} \ ,
\qquad
T'=  {\bar S' (\hat
p'^3 - \hat
p'^1)  \over  
(\hat q'^{0} -
\hat q'^{2}) } \ ,
\qquad
U'=  {\bar S' (-\hat
p'^3 - \hat
p'^1)  \over 
(\hat q'^{0} -
\hat q'^{2}) }   \ .
\end{equation}
Interestingly, we may look at the black hole entropy for both BPS and non-BPS cases, starting from the expression for the simple, non-axion case, with only 4 charges out of 8 non-vanishing and generalize it to the case when all 8 charges are present, using the $SL(2,\mathbb{Z})$ symmetry. 
The manifestly $SL(2,\mathbb{Z})\times SO(2,2)$-invariant entropy of the BPS black holes is conveniently described by the determinant of the matrix
\be
Q_{BPS}=
\begin{pmatrix}
 \hat p^2 & -\hat p\cdot \hat q\\
 -\hat p\cdot \hat q & \hat q^2 
\end{pmatrix}\ , \qquad {S\over \pi} = {A\over 4 \pi}= \sqrt {\det Q_{BPS}} \ .
\ee
Thus the entropy is manifestly $SL(2,\mathbb{Z})$ invariant, and of course, each scalar product in $Q$ is manifestly $SO(2,2)$ invariant.
The entropy of the non-BPS black holes is in turn described by the minus determinant of the matrix
\be
Q_{nonBPS}=
\begin{pmatrix}
 (\hat p')^2 & -(\hat p\cdot \hat q)'\\
 -(\hat p\cdot \hat q)' & (\hat q')^2
\end{pmatrix}\ , \qquad {S\over \pi} = {A\over 4 \pi}= \sqrt {-\det Q_{nonBPS}} \ .
\ee
The entropy formula which is manifestly $SL(2,\mathbb{Z})\times SO(2,2)$ invariant in both cases is given by 
\be
{S\over \pi} = {A\over 4 \pi}= \sqrt {|\det Q|} \ .
\ee
One can also switch to a basis in which the STU black holes have  $(SL(2,\mathbb{Z}))^3$ symmetry and where the entropy/area formula \cite{Behrndt:1996hu} is given by the Caley's determinant  of the $2\times 2 \times 2$ matrix $a_{ijk}, i,j,k =1,2$, as shown in \cite{Duff:2006uz}, \cite{Kallosh:2006zs}. The resulting entropy formula will be expressed via the set of charges $(p,q)$ and will be given in the BPS case by the 
\be
\left ({S\over \pi}\right)_{BPS} = {A\over 4 \pi}= \sqrt {-\rm Det~ a_{ijk}}
\ee
and by 
\be
\left({S\over \pi}\right)_{nonBPS} = {A\over 4 \pi}= \sqrt {\rm Det~ a_{ijk}} \ .
\ee
For both cases the entropy is
\be
{S\over \pi} = {A\over 4 \pi}= \sqrt {|\rm Det~ a_{ijk}|} \ .
\ee

\subsection{$E_{7(7)}$ black holes}

It has been shown in \cite{Kallosh:1996uy} that the entropy of the regular black holes in N=8 supergravity is given by the $E_{7(7)}$-invariant formula depending on Cartan-Cremmer-Julia quartic invariant \cite{Cartan} $J_4$. For BPS case
\be
{S\over \pi} = {A\over 4 \pi}= \sqrt {-J_4} \ ,
\label{E77}\ee
 since $J_4$ is negative.
The classification of the BPS black holes in N=8 supergravity  was performed in \cite{FM}. It was shown  in \cite{GN} that the quartic $E_{7(7)}$-invariant $J_4$ can be  positive as well as negative. Finally, the relation between the $E_{7(7)}$ black holes  and the STU black holes  was explained in 
\cite{Kallosh:2006zs}. 

One can relate the BPS to non-BPS black holes in N=8 supergravity via canonical transformation  as follows. First of all one can reduce the $E_{7(7)}$ case to the STU case and use the arguments in the previous section. Alternatively, one can proceed as follows. The Cartan-Cremmer-Julia form of the invariant \cite{Cartan}  depends on the central charge matrix $Z$,
\begin{eqnarray}
J_{4} &=& + {\mbox Tr} (Z\bar Z )^2 -{\textstyle{1\over 4}}
({\mbox Tr}\,Z\bar Z )^2 + 4  ({\mbox P\hskip- .1cm f}\; Z +
{\mbox P\hskip- .1cm f} \; \bar Z \,)  \ ,
\label{diamond}
\end{eqnarray}
or on the quantized charge matrix $(x, y)$
\begin{eqnarray}
J_{4}&=&-{\mbox Tr} (\, x \; y )^2 + {\textstyle{1\over 4}}
({\mbox Tr}\, x \;   y)^2 - 4  ({\mbox P\hskip- .1cm f}~ x +
{\mbox P\hskip- .1cm f} ~ y \,)  \ .
\label{cartan}\end{eqnarray}
Here
\begin{equation}
Z_{AB} = -{1\over 4\sqrt 2}(x^{ab} + i y_{ab})(\Gamma^{ab})_{AB}
\label{Z}\end{equation}
is the central charge matrix and
\begin{equation}
x^{ab} + i y_{ab} = -{\sqrt 2\over 4} Z_{AB}  (\Gamma^{AB})_{ab}
\label{xy}\end{equation}
is a matrix of the quantized charges related to some numbers of branes.
The exact relation between the Cremmer-Julia  invariant  in eq. (\ref{diamond}) and the Cartan invariant  in eq. (\ref{cartan})  and  has been established in \cite{Balasubramanian:1997az} and in \cite{GN}. In BPS case $J_4$ is negative.
Consider the quartic invariant in the canonical basis \cite{Kallosh:2006zs}:
\be
(x^{ab} + i y_{ab})_{\rm can}= 
\begin{pmatrix}
  \lambda_{1} & 0 & 0 & 0  \\
 0 & \lambda_{2} & 0 & 0  \\
 0 & 0 & \lambda_{3} & 0  \\ 0 & 0 & 0 & \lambda_{4} \\
\end{pmatrix}\otimes \begin{pmatrix}
 0 & \rm I \\
 -\rm I & 0 
\end{pmatrix}
\label{Zcan}\ee
The canonical transformations can be performed in this basis
\be \label{canE77}
\begin{pmatrix}
 y \\
x 
\end{pmatrix}'= 
\begin{pmatrix}
  A & B \\
  C & D 
\end{pmatrix}
\begin{pmatrix}
 y \\
x
\end{pmatrix} \ , \qquad \begin{pmatrix}
  A & B \\
  C & D 
\end{pmatrix} \subset Sp(8, \mathbb{Z}) \ .
\ee
We may now start with a particular supersymmetric configuration where all $y_{ab}$ vanish and all the off-diagonal elements of $x^{ab}$  are present. The quartic invariant is now reduced to 
\begin{eqnarray}
J_{4}&=& - 4 \,  {\mbox P\hskip- .1cm f }~ x = -4 \, x^{12} x^{34} x^{56} x^{78} <  0  \ .
\label{cartan1}\end{eqnarray}
It must be negative for BPS black holes, the entropy is
\begin{eqnarray}
{S\over \pi}&=& {2\over \pi} \sqrt{ \, x^{12} x^{34} x^{56} x^{78} } = {2\over \pi} \sqrt{ -J_4  } \ .
\label{entr}\end{eqnarray}
Now we can make a canonical transformation (\ref{canE77}) where
\be\label{canon}
A= D=
\begin{pmatrix}
  -1~ & ~0~  & ~0~ & ~0~ \\
  0 & 1 & 0 & 0 \\
   0 & 0 & 1 & 0 \\
  0 & 0 & 0 & 1 
\end{pmatrix} \ , \qquad B=C=0 \ .
\ee
The new charge vector $(x, y)'$ is now canonically related to an old charge vector $(x, y)$ as follows 
\be 
(x, y)'=  (0, 0, 0, 0, x^{12}, x^{34}, x^{56}, x^{78})'= (0, 0, 0, 0, -x^{12}, x^{34}, x^{56}, x^{78}) \ .
 \ee
The quartic invariant in terms of new charges is
\begin{eqnarray}
J_{4}&=&  - 4 \,  {\mbox P\hskip- .1cm f }~ x' = -4 \, (x^{12})' (x^{34})' (x^{56})' (x^{78})' >  0  \ ,
\label{cartan1}\end{eqnarray}
it is positive and it represents the non-BPS black hole entropy
\be
{S\over \pi} = {A\over 4 \pi}= \sqrt {J_4} \ .
\ee
Thus the entropy of the most general extremal BPS and non-BPS black holes in $N = 8$ supergravity
is given by 
\be
{S\over \pi} = {A\over 4 \pi}= \sqrt {|J_4|}
\label{E77}\ee 
 and the BPS case $J_4<0$ is related to the non-BPS case $J_4>0$
 by a  canonical  transformation.

\section{Discussion}

The fact that extremal BPS and non-BPS black holes are related by a canonical transformation is not surprising.  In both cases we are solving equations of motion of supergravity, so the non-supersymmetric solution has a spontaneously broken supersymmetry. As explained in \cite{Bellucci:2006ew}, the new attractor equations for the non-supersymmetric extremal black holes \cite{Kallosh:2005ax} correspond to a vanishing of the determinant of the fermionic mass matrix, specific for spontaneous breaking of supersymmetry. 

We have shown in this note that the one-dimensional system describing the evolution towards the horizon of the double-extremal black holes is an integrable system in ``action-angle'' variables. This means that  all possible solutions, supersymmetric or not,  should be generated by  canonical transformations. Indeed we have found few examples of the BPS and non-BPS  double-extremal black holes related by canonical transformations.

\ 
  
%{\Large {\bf Acknowledgments}}

I am grateful to A. Giryavets, A. Linde, N.
Sivanandam and  M. Soroush 
for useful conversations. This work was supported by
NSF grant PHY-0244728.

\end{document}